\documentclass[intlimits,twoside,a4paper]{article}
\usepackage[cp1251]{inputenc}

\usepackage[eqsecnum]{cmpj3}

\issue{2021}{24}{3}{33502}
\doinumber{10.5488/CMP.24.33502}
\title[Electrostatically assisted macroion association]
{Electrostatically assisted macroion association} 

\author[J. Re\v{s}\v{c}i\v{c}]{J. Re\v{s}\v{c}i\v{c} \orcid{0000-0003-0590-305X}\thanks{Corresponding author: \email{jurij.rescic@fkkt.uni-lj.si}.} }
\address{Faculty of Chemistry and Chemical Technology, University of Ljubljana, Slovenia}
\date{Received June 08, 2021, in final form July 15, 2021}

\begin{document}

\maketitle

\begin{abstract}
A model system of highly asymmetric polyelectrolyte with directional short-range 
attractive interactions was studied by canonical Monte Carlo computer simulations. 
Comparison of MC data with previously published theoretical results shows good
agreement. For moderate values of binding energies, which matches those of molecular 
docking, a dynamic equilibrium between free and dimerized macroions is observed. 
Fraction of dimerized macroions depends on macroion concentration, binding energy 
magnitude, and on the valency of small counterions. Divalent counterions induce an 
effective attraction between macroions and enhance dimerization. This effect is 
most notable at low to moderate macroion concentrations.  
\keywords{polyelectrolyte, association, electrostatics, Monte Carlo simulation}%
\end{abstract}

\section{Introduction}
 
Charged colloidal systems are of great scientific and technological importance, as they include
systems such as micellar solutions, metal-oxide nanoparticle suspensions, and protein solutions~\cite{wenner}. 
These systems have long been subject to extensive experimental and theoretical research. One of the most widely 
used theoretical approaches used to study the properties of charged colloidal systems is the classical 
Derjaguin-Landau-Verwey-Overbeek (DLVO) theory~\cite{dl,vo}. In contrast to this theory, which predicts purely 
repulsive forces between similarly-charged colloidal particles and was used to explain stability of colloidal systems, 
several studies presented evidence of effective attractive forces between macroions in these systems. Khan et. al.~\cite{khan} 
reported an unusual phase behaviour of surfactant systems with divalent counterions. Kjellander and Mar\v{c}elja observed 
an attractive force between two mica surfaces in presence of divalent ions~\cite{kjellander}. Ise observed void formation
in salt-free latex suspensions~\cite{ise}.
In addition to experimental work, computer simulations were also extensively used to address these observations~\cite{lobaskin98,linse2000}.
Later, it has been confirmed that asymmetric polyelectrolytes undergo phase separation~\cite{jure2001,hynninen}.

Associating systems are particularly important, as they mimic protein association~\cite{matthews,rescic2001,rescic2016,kastelic16}, 
enzyme-ligand binding, and molecular docking. Many proteins are prone to dimerization or oligomerization~\cite{alm,dip}; this 
process often increases protein stability or grants them additional functionality. In these systems, directional short-ranged 
interactions are at work along with other intermolecular forces. Structural and thermodynamic properties of liquids are successfully 
described with theories based on the Ornstein-Zernike integral equation. Associative hypernetted chain approach, developed by 
Kalyuzhnyi and co-workers~\cite{ykvv93,ykvv95}, was further extended with Wertheim's~\cite{wertheim1,wertheim2} two-density 
theory for associating liquids to yield a powerful tool for studies of dimerizing charged colloidal systems~\cite{ykvv98}.

Theoretical description of associating fluids has seen additional progress over the last decade in a form of the
resummed thermodynamic perturbation theory for associating fluids with multiple bonding sites~\cite{yura2010}. 
Patchy colloids can also be treated as associating systems and have been studied with integral equation theories 
and computer simulations~\cite{marshall2013,nezbeda2020}. 
 
The aim of this work is to estimate a contribution of long-range electrostatic interactions to 
macroion association and to provide a comparison with already published theoretical data~\cite{ykvv98}.

\section{Model and method}

The salt-free polyelectrolyte solution model consists of spherical macroions with radius $R_{\rm{M}}=15$~\AA~and 
spherical counterions with radius $R_{\rm{C}}=2$~\AA. Macroion charge $Z_{\rm{M}}$ is equal to $-15e_0$ while the counterion charge 
$Z_{\rm{C}}$ is $1e_0$ or $2e_0$. 

Total potential energy of the model system is assumed to be pairwise additive and consists of three contributions:
(i) hard-sphere repulsion, (ii) electrostatic interactions, and (iii) associative interactions.

\begin{equation}
U=U_{\rm{HS}}+U_{\rm{elec}}+U_{\rm{assoc}},
\end{equation}
with $U_{\rm{HS}}=\sum_{i<j}u_{\rm{HS}}(r_{ij})$ and

\begin{equation}
u_{\rm{HS}}(r_{ij}) = \left\{ \begin{array}{ll}
\infty; & r_{ij} < R_{i} + R_{j}\\
     0; & r_{ij} \geqslant R_{i}+R_{j}.
\end{array} \right.
\end{equation}
$R_{i}$ and $R_{j}$ are radii of particles $i$ and $j$, respectively. The $r_{ij}$ is the corresponding
centre-to-centre separation.

Electrostatic term has the following form:

\begin{equation}
U_{\rm{elec}}=\sum_{i<j}u_{\rm{elec}}(r_{ij})=\sum_{i<j}\frac{Z_iZ_j e_0^2}{4 \piup \epsilon_0 \epsilon_r r_{ij}},
\end{equation}
where $Z_i$ is the valency of particle $i$, $\epsilon_0$ is the dielectric permittivity of vacuum and 
$\epsilon_r$ is the relative permittivity of water which is equal to 78.4 at a room temperature.

\begin{figure}[!t]
	\centerline{\includegraphics[width=0.50\textwidth]{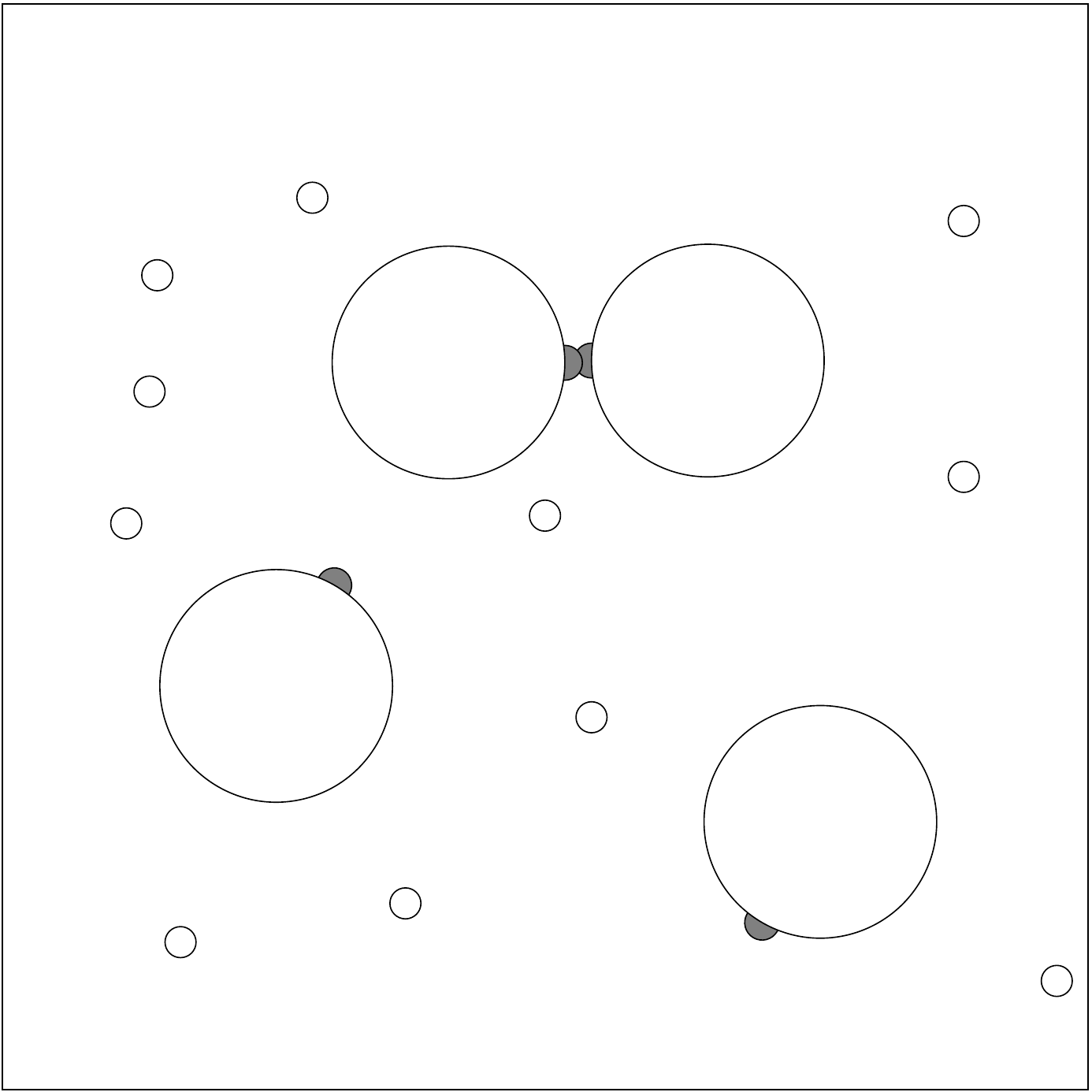}}
	\caption{Schematic representation of the model system. Associative regions are darker.
	} \label{mod1}
\end{figure}

Association is possible only between macroions, as they possess one associative site per molecule. The origin 
of an associative site is located on the surface of a macroion and the site is spherical in shape. 
The associative part of the intermolecular potential is given by:

\begin{equation}
u_{\rm{assoc}}(r_{ij}) = \left\{ \begin{array}{ll}
-\epsilon_{\rm{assoc}} & r_{\rm{site-site}} \leqslant 2\Delta\\
     0; & r_{\rm{site-site}} > 2\Delta,
\end{array} \right.
\end{equation}
$\Delta$ is the square well width and is chosen to be equal to $0.264R_{\rm{M}}$, which allows only for formation of dimers. 
Since one of the goals of this research is to provide a comparison with the existing theoretical study, the model parameters 
are the same as in the work of Kalyuzhnyi and Vlachy~\cite{ykvv98}. The model is schematically represented in  figure~\ref{mod1}. 

Linse thoroughly investigated salt-free colloidal suspensions and proposed a set of 
three parameters which describe these systems~\cite{linse2000}. These are macroion-counterion charge ratio $Z_r=|Z_{\rm{M}}/Z_{\rm{C}}|$, 
macroion volume fraction $\Phi_{\rm{M}}=4\piup R_{\rm{M}}^3 \rho/3$, and electrostatic coupling parameter $\Gamma_{II}=Z_{\rm{C}}^2L_B/R_{\rm{M}}$, 
with $L_B=\beta e_0^2/(4\piup\epsilon\epsilon_0)$ being the Bjerrum length and $\beta=1/(k_{\rm{B}}T)$. 
Large values of $\Gamma_{II}$ indicate phase transition of a colloidal system.

\subsection{Monte Carlo simulations}
 
The model system was used with the canonical ensemble Monte Carlo computer simulation following the standard Metropolis sampling scheme. 
A simulation box was a cube with dimensions which correspond to a given macroion concentration. The number of macroions $N_{\rm{M}}$ was 64 while the 
number of counterions was 
$N_{\rm{C}}=N_{\rm{M}} \times |Z_{\rm{M}}/Z_{\rm{C}}|$
 to fulfill electroneutrality. Macroions were initially placed on a 3D grid while counterions 
were inserted randomly. For a production run, at least 1 million moves per particle were performed. Electrostatic interactions were 
calculated using Ewald summation. Along with the excess internal energy, radial distribution functions, and cluster size distributions were 
calculated. Moreover, the evidence of bonds (which macroions are bonded) was maintained and updated at every macroion move. A fraction of bonded 
macroions was calculated from the collected bond histogram. All simulations were performed at a temperature of 298~K. Several square well depth 
values were studied: $\beta \epsilon_{\rm{assoc}}$ = 5, 8, 10, 20, 25, 33, and 43. Each $\beta \epsilon_{\rm{assoc}}$ value can represent a different 
enzyme since each one has a unique binding energy for a matching ligand. Macroion concentration  varied from 0.004~M to 0.05~M. Highly charged 
macroions are tightly surrounded by counterions, resulting in a very high rejection rate of macroion single-particle trial moves.
To overcome this issue, cluster moves were applied. Typically, displacement parameters are adjusted to the values which yields 50\% acceptance rate. 
This is not always the best strategy to achieve a fast convergence and some techniques have been proposed for choosing optimal displacement 
parameters~\cite{hebb2016}. For highly charged systems, a total r.m.s. displacement for each particle type is monitored and should be of the size 
of a simulation box or larger. Associated particles should be allowed to explore various bonded configurations~\cite{rescic2016}. All these 
approaches are included in the MOLSIM simulation package~\cite{molsim} which was used to perform simulations.

\section{Results and discussion}

\begin{figure}[!t]
\centerline{\includegraphics[width=0.55\textwidth]{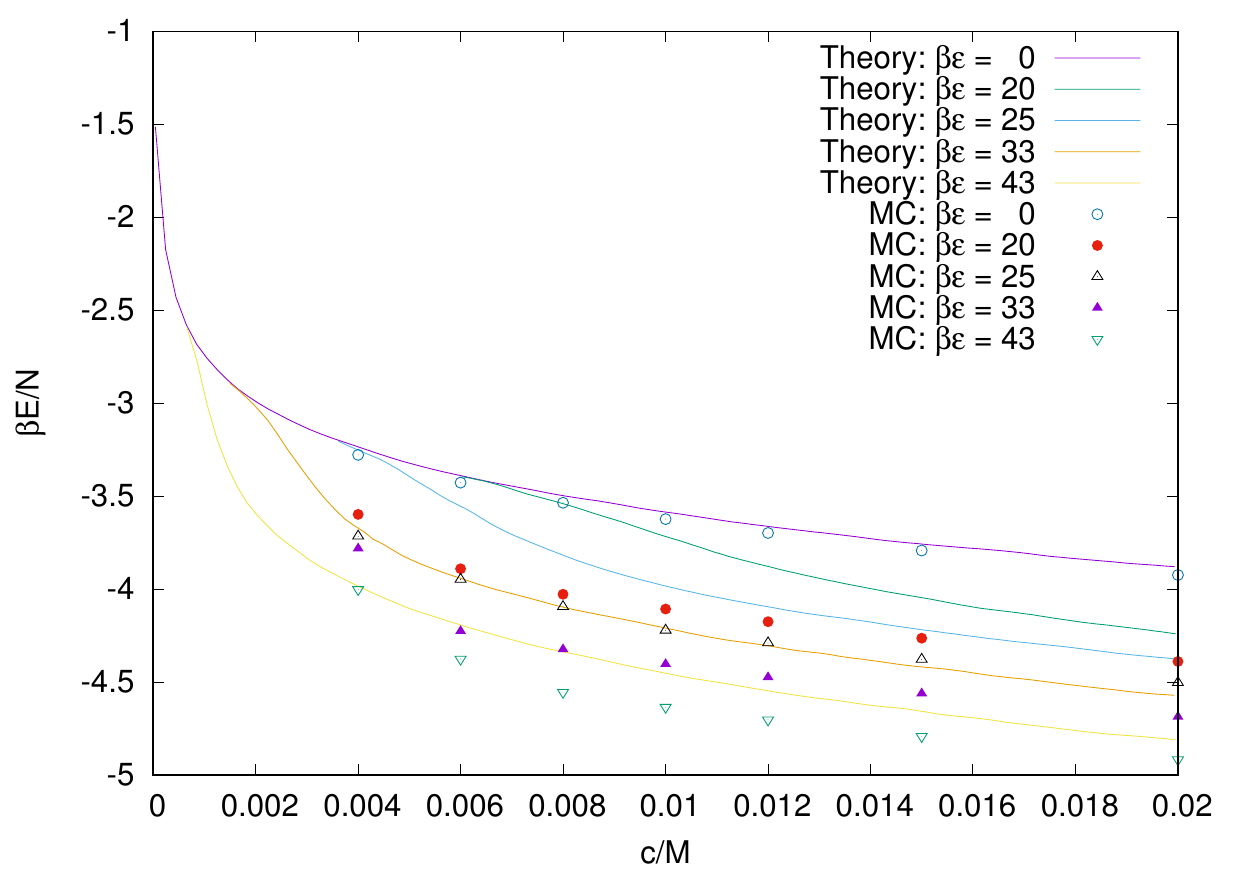}}
	\caption{(Colour online) Comparison between Monte Carlo and theoretical results. The latter are retrieved from the reference~\cite{ykvv98}.
} \label{mod2}
\end{figure}

The same model was treated via the AHNC/Wertheim formalism~\cite{ykvv98}. At first, several simulations have been carried out to compare 
theoretical results with MC data, which is displayed in  figure~\ref{mod2}. For a system without associative potential the agreement is very good. 
It must be noted that the values of associative parameter $\epsilon_{\rm{assoc}}$ of 20 $k_{\rm{B}}T$ or higher result in almost unbreakable bond --- once a dimer
is formed, both macroions forming it stay bonded.
Energies are therefore simply shifted to more negative values by the same amount for all concentrations studied at each value of the $\epsilon_{\rm{assoc}}$.
Concentrations lower than 0.004~M were not studied due to a slow convergence. 

\begin{figure}[!t]
\centerline{\includegraphics[width=0.55\textwidth]{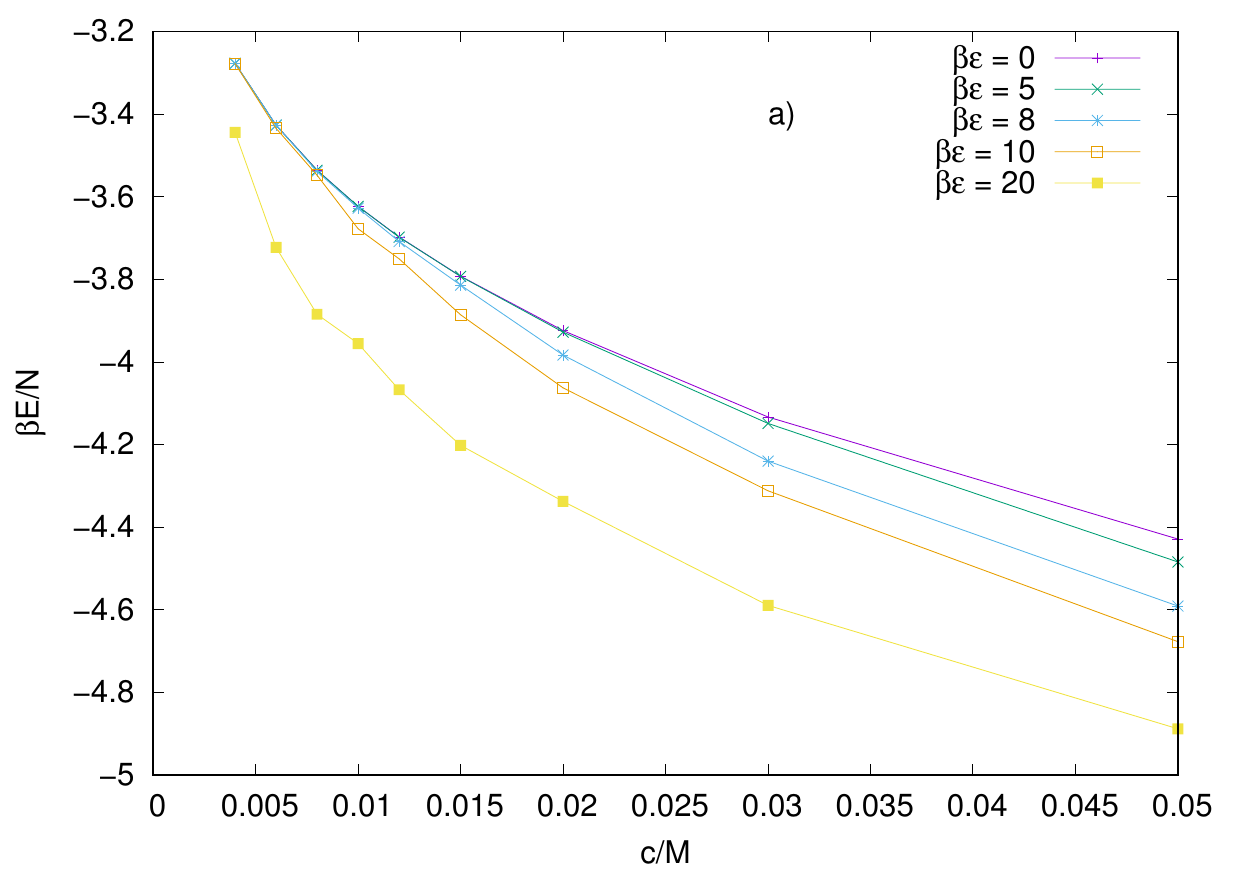}}
\centerline{\includegraphics[width=0.55\textwidth]{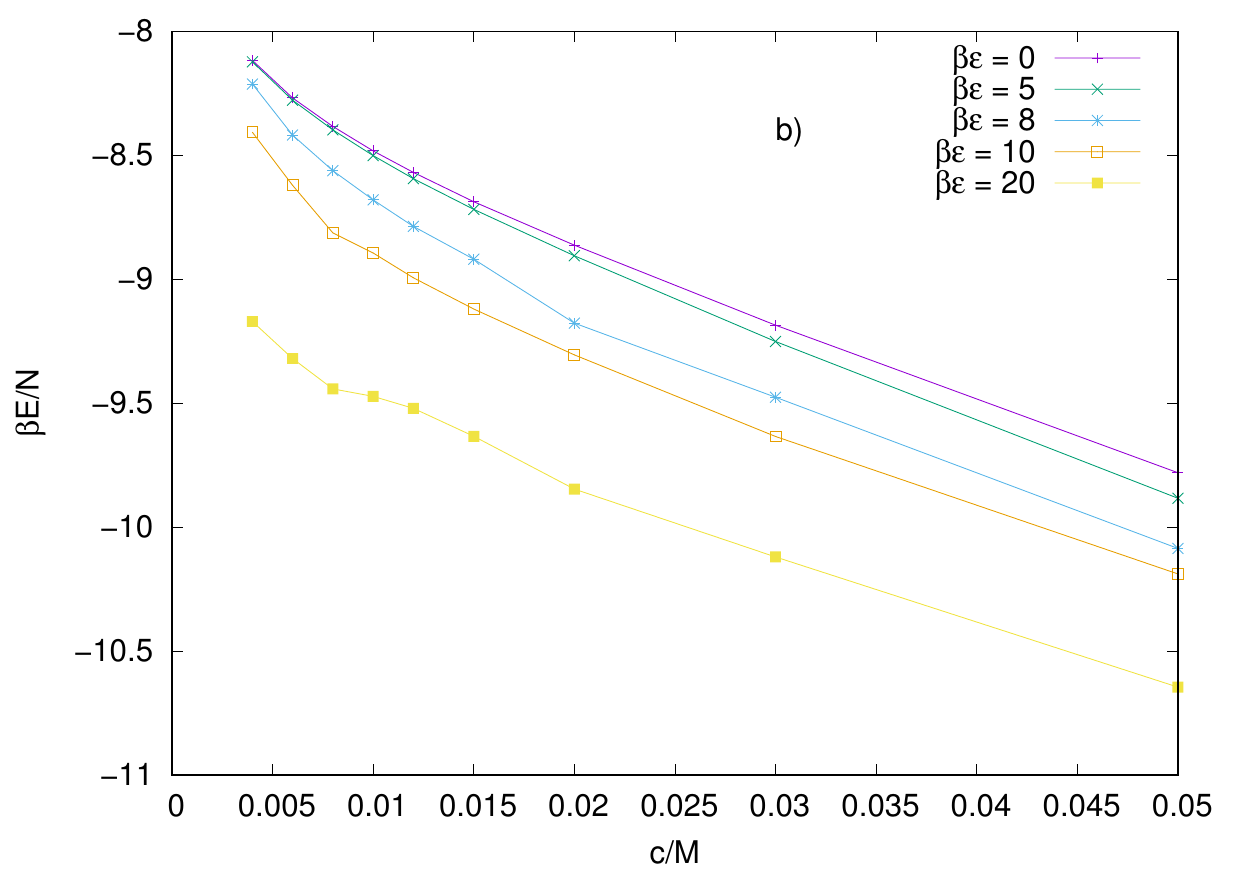}}
	\caption{(Colour online) Excess internal energies as a function of macroion molar concentration for monovalent (panel a) and 
	divalent (panel b) counterions. Associative energy is given in the legend.
} \label{mod3}
\end{figure}

Realistic enzyme/ligand binding energies are smaller in magnitude and are in a range of ~$10 k_{\rm{B}}T$. Values of $\beta \epsilon_{\rm{assoc}}$ = 5, 8, and 10 were 
used throughout the rest of the study. Excess internal energies are presented in  figure~\ref{mod3} and cover polyelectrolyte concentrations in the range from 0.004~M 
to 0.05~M. Weaker association potential allows for dynamic creation, destruction, and rearrangement of bonds between macroions. Excess internal energies are 
at lower polyelectrolyte concentrations similar to each other. At larger concentrations, more macroions are bonded which is reflected in lower energies than 
a system without associative part of the intermolecular potential. Panel a) of figure~\ref{mod3} applies to monovalent counterions. Divalent counterions induce a weak effective  
attraction between macroions, which results in a lower excess internal energy of the system. This is presented in 
the panel b) of figure~\ref{mod3}. 

\begin{figure}[!t]
\centerline{\includegraphics[width=0.55\textwidth]{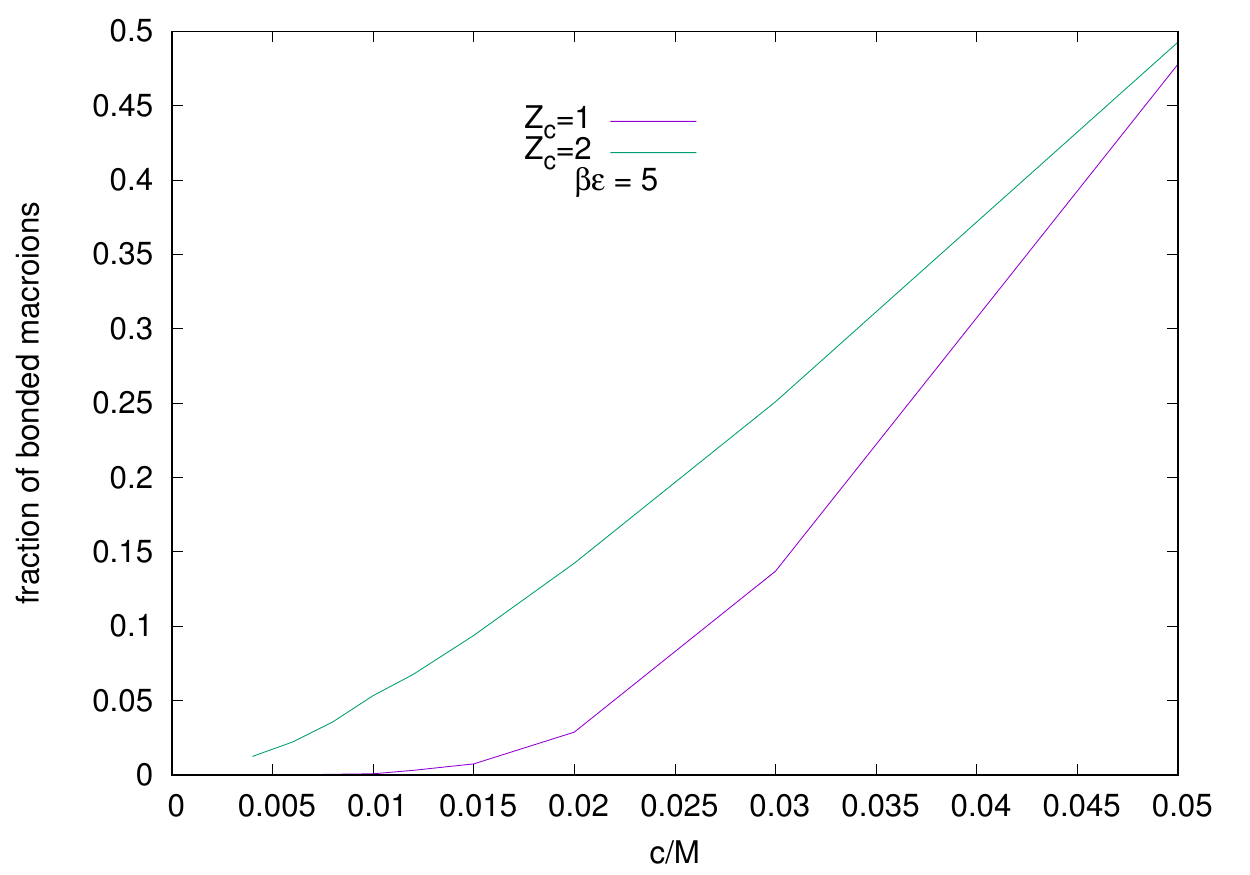}}
\centerline{\includegraphics[width=0.55\textwidth]{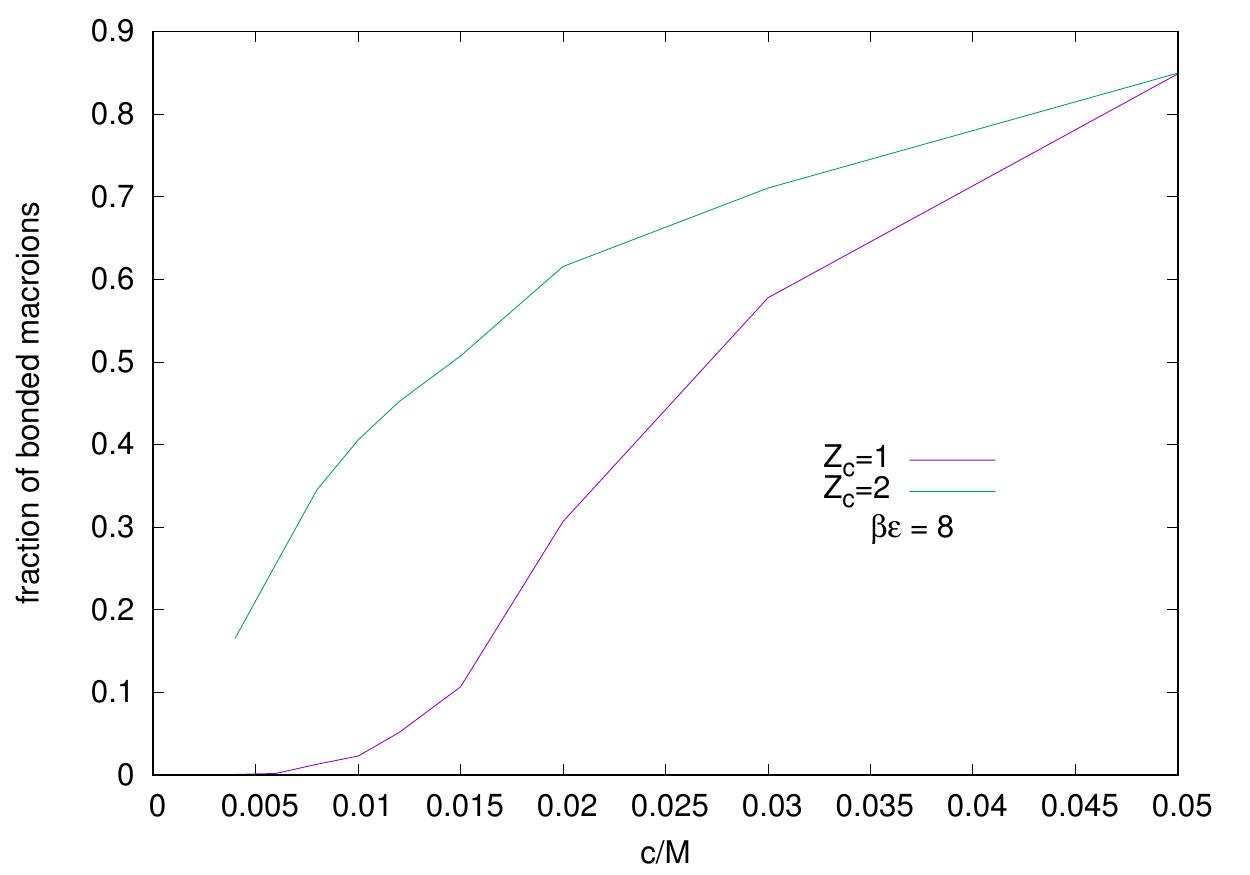}}
\centerline{\includegraphics[width=0.55\textwidth]{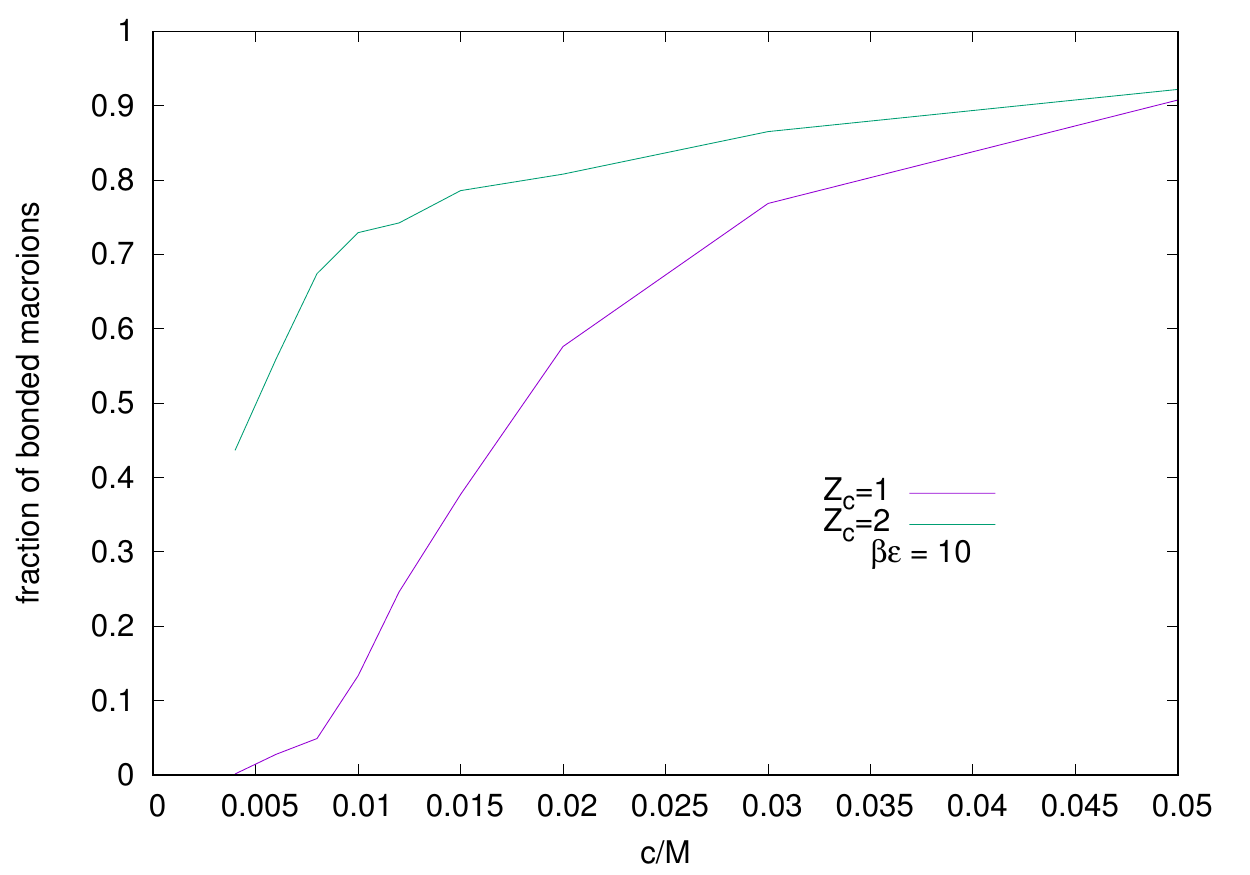}}
	\caption{(Colour online) Fraction of dimerized macroions as a function of polyelectrolyte molar concentration for three different associating energies: 
	$5k_{\rm{B}}T$ (top), $8k_{\rm{B}}T$ (middle), and $10k_{\rm{B}}T$ (bottom) for both monovalent and divalent counterions. Associative energy and counterion 
	valency is given in the legend.
} \label{mod4}
\end{figure}

Fraction of bonded macroions was calculated in all cases. Figure~\ref{mod4} presents results for three square-well depth values. Larger $\epsilon_{\rm{assoc}}$ yields a
larger fraction of dimers for all concentrations studied. At $\beta \epsilon_{\rm{assoc}} \geqslant 20$ all macroions are bonded. More realistic values of 
$\epsilon_{\rm{assoc}}$ yield lower fractions of dimers which are also concentration dependent. Electrostatic energy between bare macroions is 
repulsive and equal to $95.3k_{\rm{B}}T$ when two macroions are in a contact. This repulsion can easily overcome the associative energy which is evident for 
the smallest $\beta \epsilon_{\rm{assoc}}=5$. 
Increasing the counterion valency from 1 to 2 has a profound effect. It is known that increasing electrostatic coupling destabilizes a colloidal suspension.

The $\Gamma_{II}$ parameter, which is a measure for the electrostatic coupling, is for the model system with monovalent counterions equal to 0.476 and 
rises fourfold to 1.9 when divalent ions are present. Weak to moderate effective attractive forces start to appear among macroions. 
This in turn increases the probability of overlapping the associative regions on macroions. A similar model system without associative potential undergoes 
the gas-liquid phase separation and has a critical point at $\Gamma_{II}$ value of 2.6~\cite{jure2001}.
Presence of divalent counterions leads to a larger fraction of bonded macroions regardless of the $\epsilon_{\rm{assoc}}$ value. Both electrostatics and 
association lower the energy of the system, which is displayed in figure~\ref{mod3}. 
This effect is even more pronounced at low to moderate macroion concentration, where the increase in the fraction of dimerized macroions is nearly 10-fold. 

\begin{figure}[!t]
\centerline{\includegraphics[width=0.55\textwidth]{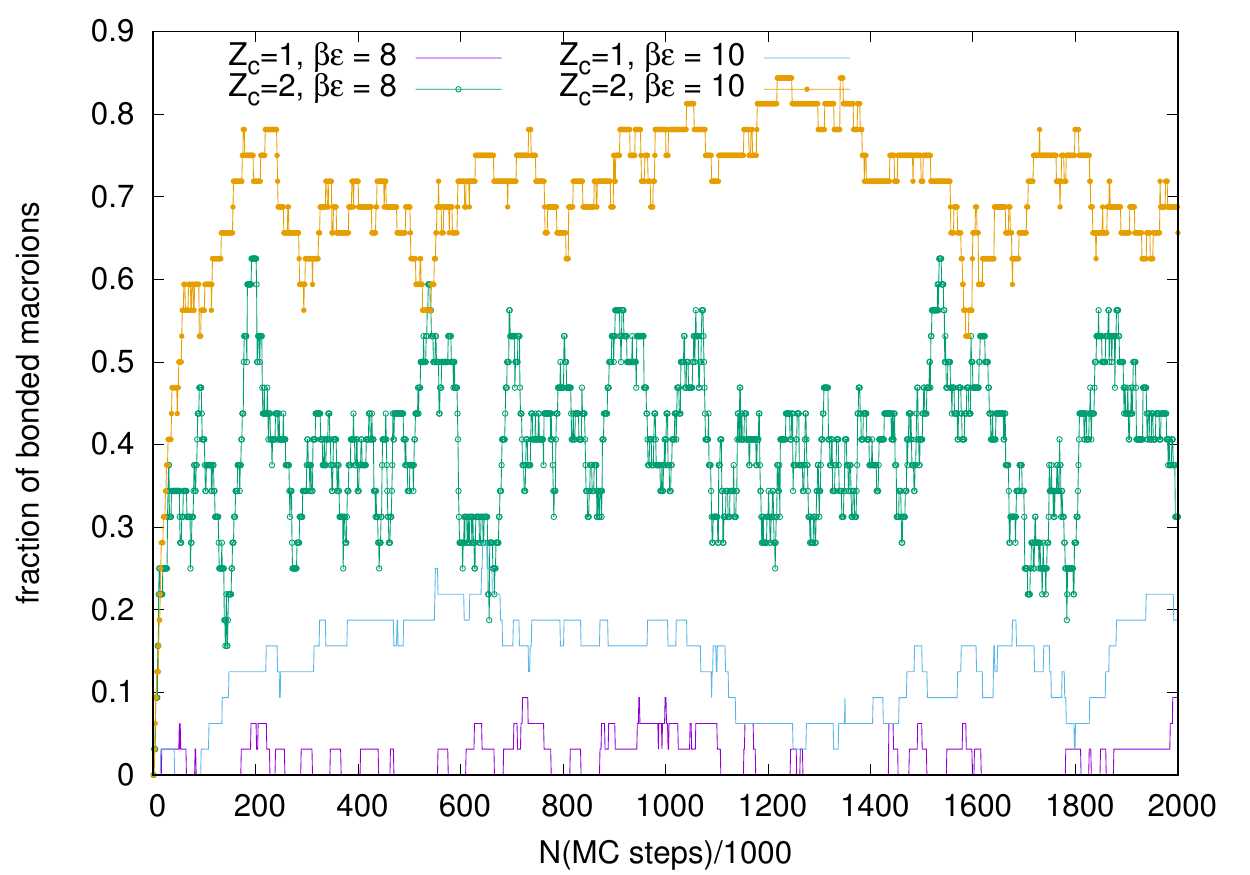}}
\centerline{\includegraphics[width=0.55\textwidth]{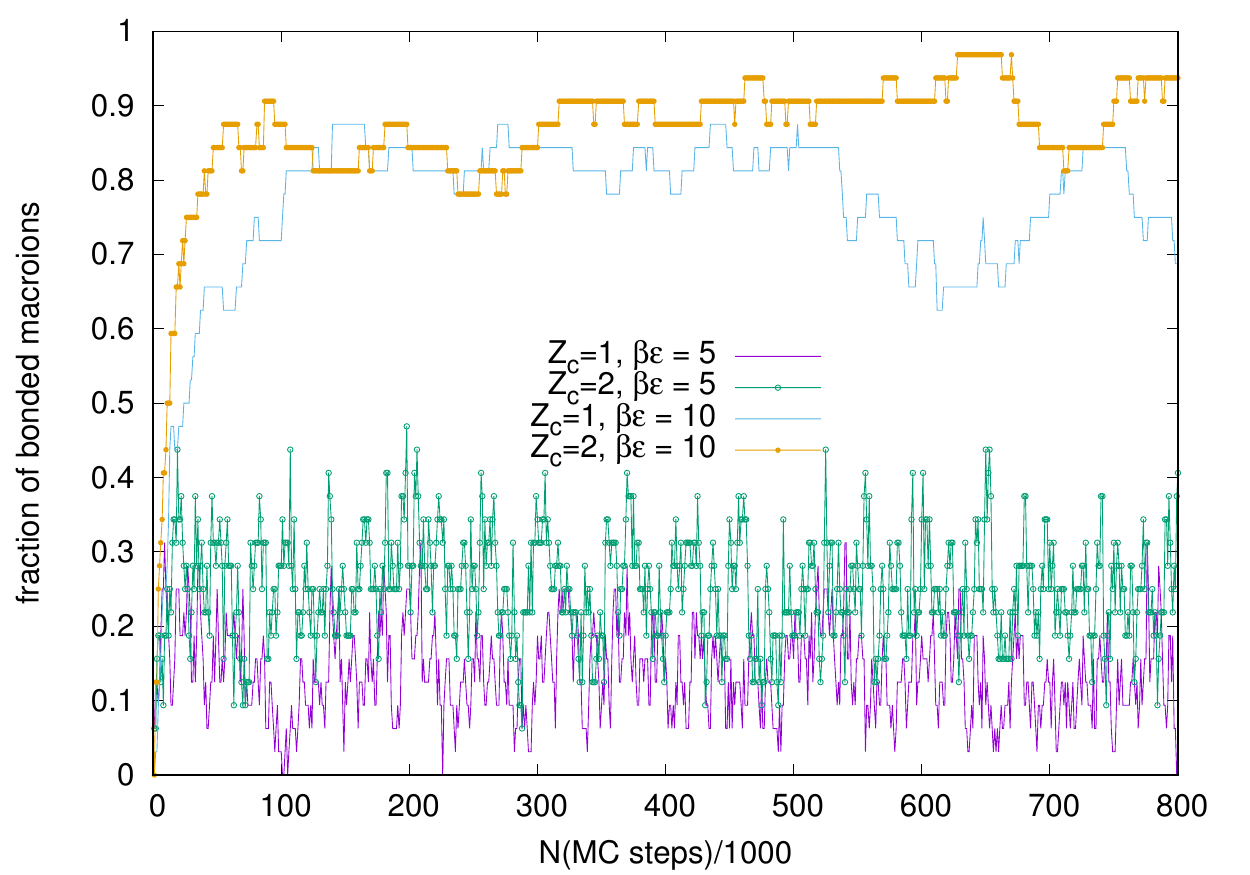}}
	\caption{(Colour online) Histogram of a fraction of dimerized macroions for two polyelectrolyte molar concentrations: 0.01~M (top) and 0.03~M (bottom) at 
	different association energies: ($5k_{\rm{B}}T$), ($8k_{\rm{B}}T$), and ($10k_{\rm{B}}T$)  for both monovalent and divalent counterions. Associative energy 
	and counterion valency is given in the legend.
} \label{mod5}
\end{figure}

Figure~\ref{mod5} provides a more detailed insight into the dimerization process dynamics. Two macroion concentrations were studied: 0.01~M (upper panel), 
and 0.03~M (bottom panel). Each panel contains information about dimerization dynamics for two values of $\epsilon_{\rm{assoc}}$ and for both monovalent 
and divalent counterions. In all cases, systems with divalent counterions contain a larger fraction of dimerized macroions, while this fraction is 
the smallest in systems with monovalent counterions and smaller $\epsilon_{\rm{assoc}}$. Larger value of $\epsilon_{\rm{assoc}}$ increases longevity of 
bonds. Systems with divalent counterions also reach the equilibrium fraction of bonded macroions faster than their monovalent counterparts. 

\begin{figure}[!t]
\centerline{\includegraphics[width=0.55\textwidth]{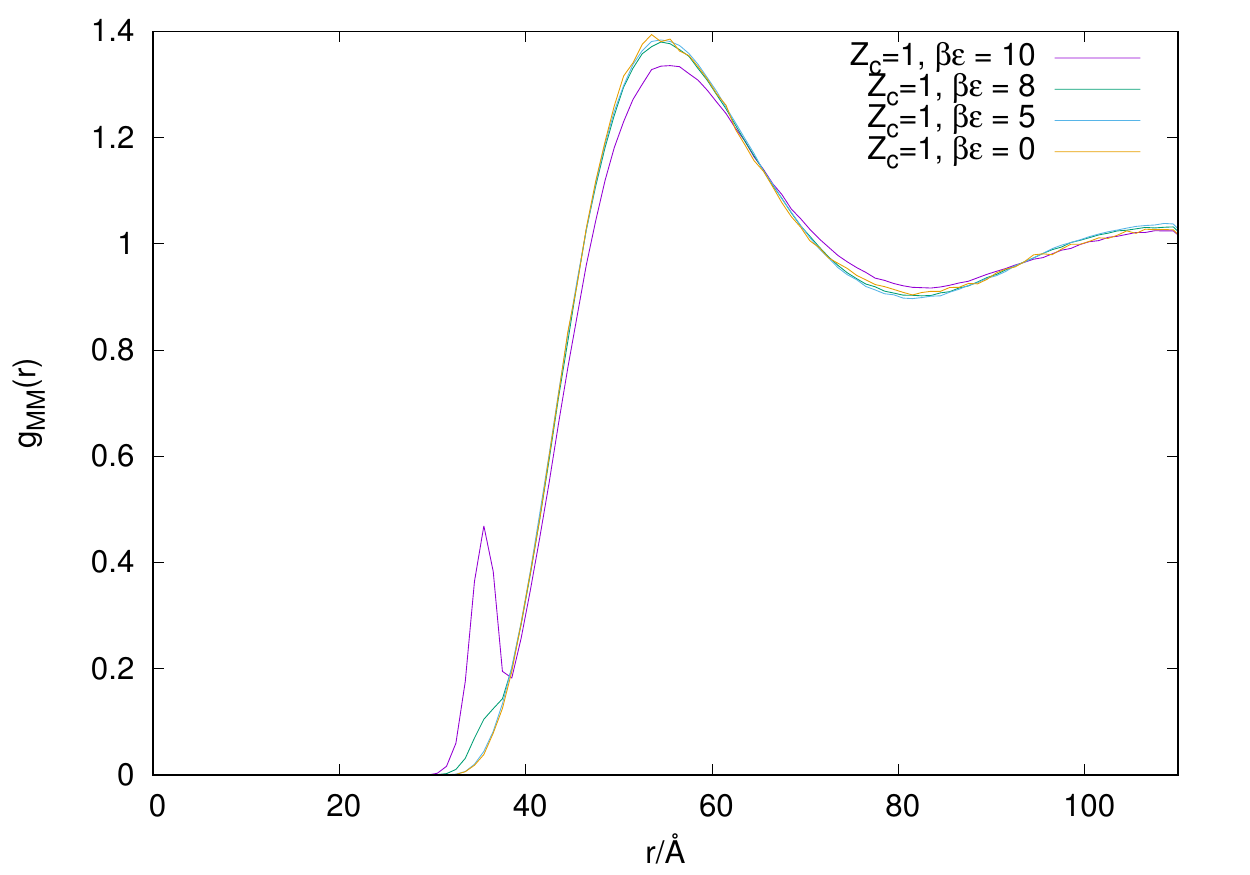}}
\centerline{\includegraphics[width=0.55\textwidth]{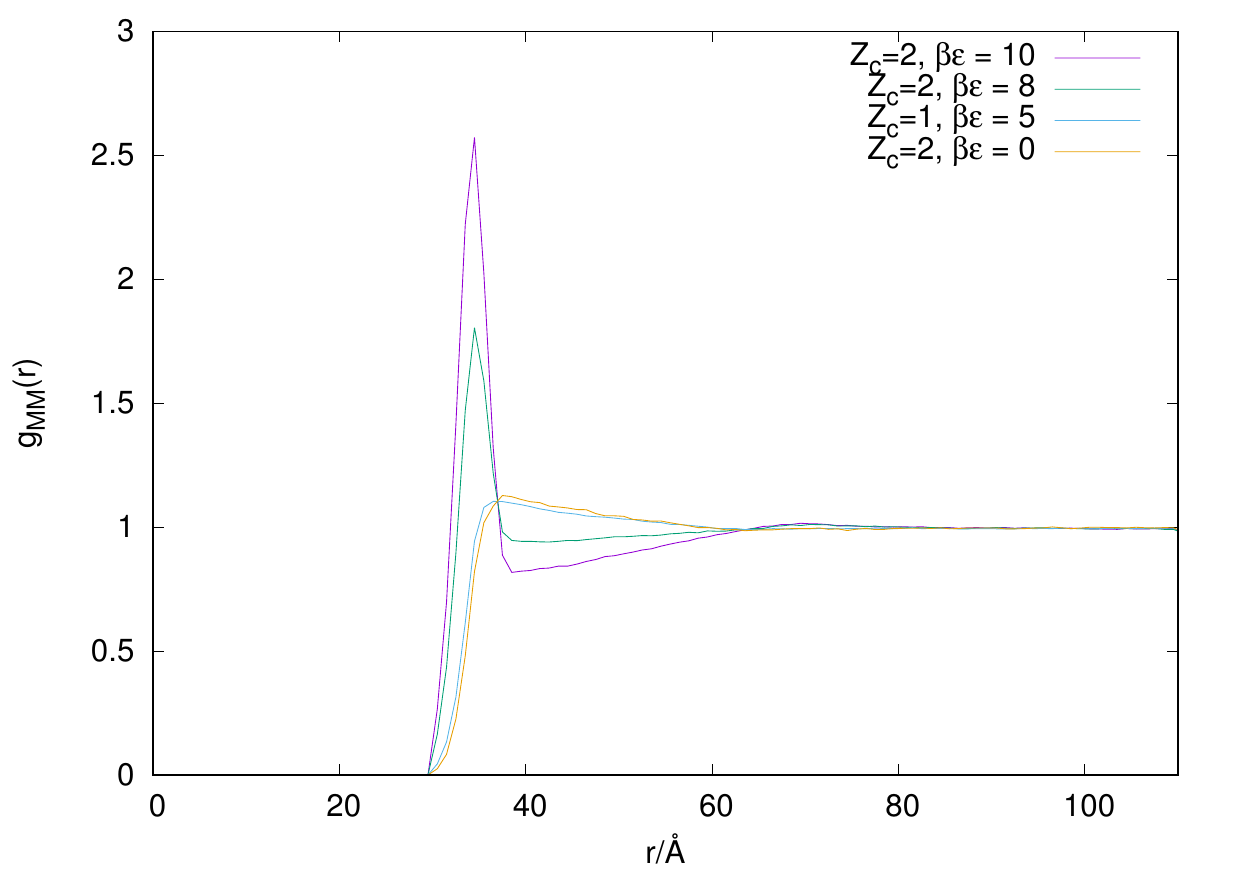}}
	\caption{(Colour online) Macroion-macroion radial distribution function for two values of associating energy and for monovalent (top) and divalent 
	couterions (bottom panel), all at macroion molar concentration $c_{\rm{M}}= 0.01$~M.
} \label{mod6}
\end{figure}

To better illustrate the effect of increased electrostatic coupling on macroion association when switching from monovalent to divalent counterions, 
macroion-macroion radial distributions $g_{MM}(r)$ are plotted in figure~\ref{mod6}. The upper panel presents $g_{MM}(r)$ in systems with monovalent counterions 
and different $\epsilon_{\rm{assoc}}$. Only at the strongest association energy of 10$k_{\rm{B}}T$, the $g_{MM}(r)$ differs qualitatively from the rest, as a small 
peak appears at the macroion-macroion distance of~36~\AA.  

Association becomes much more pronounced in systems with divalent counterions. Slight effective macroion-macroion attraction is deduced from 
the $g_{MM}(r)$ for a system without the associative potential as the broad peak around 56~\AA~disappears and a small peak appears at 38~\AA. 
Electrostatic forces bring macroions close enough so that short range associative potential can take dimerization over. Smaller average macroion-macroion 
distance increases the probability of dimerization. The peak at 36~\AA \- in $g_{MM}(r)$ is much larger and is also visible  at smaller values of $\epsilon_{\rm{assoc}}$. 
\newpage

\section{Conclusions}

A salt-free colloidal solution model was studied using the canonical Monte Carlo simulations. Each highly charged macroion possesses one site with 
short-range associative potential. Such a model is suitable to study the protein dimerization in aqueous solutions with focus on the effect of 
electrostatic forces on macroion-macroion dimerization. First, we compared the excess internal energies obtained from simulations with theoretical 
results available in the literature, finding a good agreement between the two approaches. Structures of the solutions in a form of radial distribution 
functions were also very similar. Since divalent counterions induce a weak effective attraction among macroions, it was of particular interest to 
evaluate the effect of long-range electrostatic forces on macroion dimerization. Much larger fractions of dimerized macroions were found 
even at low macroion concentration in presence of divalent counterions, showing the effect of increased electrostatic coupling. 

\section{Acknowledgement} 

Author acknowledges Slovenian Research Agency for financial support through the research program P1-0201.

\ukrainianpart

\title{Макроіонна асоціація за сприяння електростатики} 

\author[Ю. Решчіч]{Ю. Решчіч  }
\address{Факультет хімії та хімічних технологій університету Любляни, Любляна, Словенія}

\makeukrtitle
	
	\begin{abstract}
			Модельну систему сильно асиметричного поліелектроліту з направленими короткосяжними притягальними взаємодіями досліджено за допомогою канонічного моделювання Монте Карло (МК). Дані моделювання МК добре узгоджуються з попередньо отриманими теоретичними результатами. Для помірних величин енергії зв'язку, що співпадають зі значеннями молекулярного докінгу, спостерігається динамічна рівновага між вільними та димеризованими макроіонами. Частка димеризованих макроіонів залежить від їх концентрації, величини енергії зв'язку та валентності малих контріонів. Двовалентні контріони індукують ефективне притягання між макроіонами та підсилюють димеризацію. Цей ефект найбільш помітний при низьких та середніх концентраціях макроіонів.
		
				\keywords поліелектроліти, асоціація, електростатика, моделювання Монте Карло
		
	\end{abstract}

\lastpage
\end{document}